# Enhanced thermal stability of inverted perovskite solar cells by bulky passivation with pyridine-functionalized triphenylamine

*Ekaterina A. Ilicheva[1,§], Irina A. Chuyko[1,2,§], Lev O. Luchnikov[1], Polina K. Sukhorukova[1,2], Nikita S. Saratovsky[2], Anton A. Vasilev[3], Luiza Alexanyan[3], Anna A. Zarudnyaya[1], Dmitri Yu. Dorofeev[1], Sergey S. Kozlov[4], Andrey P. Morozov[1], Dmitry S. Muratov[5], Yuriy N. Luponosov[2,*], and Danila S. Saranin[1,*]*

[1]LASE – Laboratory of Advanced Solar Energy, NUST MISIS, 119049 Moscow, Russia

[2]Enikolopov Institute of Synthetic Polymeric Materials of the Russian Academy of Sciences (ISPM RAS), Profsoyuznaya St. 70, Moscow, 117393, Russia

[3]Department of semiconductor electronics and device physics, NUST MISIS, 119049 Moscow, Russia

[4]Laboratory of Solar Photoconverters, Emanuel Institute of Biochemical Physics, Russian Academy of Sciences, 119334 Moscow, Russia

[5]Department of Chemistry, University of Turin, 10125, Turin, Italy

[§]The authors contributed equally to this work

**Corresponding authors:**

Dr. Yu. N. Luponosov luponosov@ispm.ru,

Dr. Danila S. Saranin saranin.ds@misis.ru.

**Funding:** Russian Science Foundation (RSF) (grant № 22-19-00812-P)

**Key-words:** halide perovskites, self-assembling monolayers, passivation, phase stability, molecular design

**Abstract**

Despite competitive efficiency compared to Si solar cells and relevant stability at near room temperatures the rapid degradation at elevated temperatures remains the critical obstacle for exploitation of perovskite photovoltaics. In this work, a 4-(pyridin-4-yl)triphenylamine (**TPA-Py**) with pyridine anchor group was employed for inter-grain bulk modification of double-cation $CsCH(NH_2)_2PbI_3$ perovskite absorbers to enhance thermal stability. Through coordination and dipole-dipole interactions, nitrogen-containing fragments (diphenylamine and pyridine) of TPA-Py passivate uncoordinated cations and improve the phase resilience of perovskite films against segregation. This resulted in a power conversion efficiency of 21.3% with a high open-circuit voltage of 1.14 V. Notable impact of self-assembled monolayer incorporated into the bulk of perovskite film manifested in huge improvement of thermal stability at 85°C (ISOS-D-2). TPA-Py modification improved extended the T80 lifetime to ~600 h compared to only 200 h for the reference under harsh heating stress in ambient conditions. In-depth analysis using photoinduced voltage transients and admittance spectroscopy after different stress periods revealed the screening of ion migration (0.45 eV) for devices with TPA-Py. This work offers an important understanding of the bulk modification of microcrystalline perovskite absorbers and guide for robust design of bulk and buried interfaces in highly efficient perovskite solar cells.

1. **Introduction**

Over the past decade, halide perovskite photovoltaics (**HP PV**s) have demonstrated unprecedented progress in achieving high power conversion efficiencies (**PCE**s) as well as in advancing solution-processing technologies[1]. By 2025, the record efficiency of laboratory-scale perovskite solar cells has reached 27.0%[2], and several technology companies have announced the launch of pilot-scale production[3]. Nevertheless, the challenge of long-term stability remains unresolved. The formation of various charged defects, such as vacancies, antisite defects, and interstitials, particularly at the interfaces of microcrystalline perovskite absorber films, induces ion migration, electrochemical corrosion, and molecular decomposition of the perovskite lattice[4]. These phenomena significantly limit the long-term operational stability of halide perovskite-based devices.

Recent literature reports demonstrate that the long-term stability of perovskite solar cells, ranging from hundreds to several thousand hours, can be achieved through various approaches, including passivation and/or compositional engineering of the perovskite lattice. However, such stabilization is often documented under relatively mild conditions, such as room temperature[5–7]. At elevated temperatures approaching realistic operating conditions of solar panels (85-100 °C), a persistent degradation trend is typically observed[5,8,9]. High-efficiency PV devices typically employ HPs with the general formula $APbI_3$, where A is an organic cation such as formamidinium (FA), methylammonium (MA), or cesium[10]. The presence of defects within the perovskite lattice leads to distortions of the crystal structure and deviations from stoichiometry, which in turn promote undesirable phase transitions. Thermal stress induces the activation of ion migration (iodine vacancies)[11], as well as uncompensated Pb-centers[12] in parallel to formation of $PbI_2$ at grain boundary interfaces[13]. The activation threshold for rapid thermal degradation in perovskite absorbers containing an organic A-cation (MA, FA) corresponds to temperatures above 85 °C[14].

One of the most promising state-of-the-art strategies for interface modification in halide perovskite photovoltaics is the use of self-assembled monolayers (SAMs)[15,16]. The formation of ultrathin SAM layers is possible due to the presence of an anchor group in their structure, which is capable of specific interactions with the substrate material or surface of the coatings. These molecules enable the creation of surface dipoles, electrostatic passivation [17], and energy band bending, thereby improving charge collection dynamics and suppressing the accumulation of ionic defects.

One of the most frequently used anchor groups is the phosphoric anchor group (-PO(OH)$_2$), which contains a phosphorus atom bonded tetrahedrally to three oxygen atoms, typically two hydroxyls (-OH) and one double-bonded oxygen (=O). This allows for multiple robust binding modes with oxide surfaces, including bidentate bridging, where two oxygen atoms bond to two metal atoms on the surface; tridentate modes, involving three oxygen atoms; and additional hydrogen bonding, which stabilizes the structure[18]. However, the phosphoric group's strong acidity could corrode reactive metal oxide surfaces such as nickel oxide and thus undermine the stability of the device[19].

Triphenylamine (TPA)-based self-assembled monolayers (SAMs) have been identified as a promising solution for perovskite solar cells, exhibiting several key advantages, including remarkable electrochemical and phase behavior, as well as thermal stability. The TPA core exhibits enhanced hole mobility and is more susceptible to oxidation and stabilization by hyperconjugation effects, thereby augmenting charge extraction efficiency. Furthermore, TPA-based SAMs demonstrate exceptional thermal and morphological stability, along with the capacity to precisely adjust energy levels, rendering them highly compatible with perovskite layers. Furthermore, TPA-core easily accessible chemical functionalization, thereby enabling the majority of structural changes required for the optimization of material properties[20,21].

Defect passivation and bonding to surfaces depend on the SAM's interactions with both the charge transport layer's surface and the perovskite's surface. Chemical interactions with under-coordinated lead atoms in the perovskite absorber can occur in the presence of nitrogen-containing anchoring groups within the molecular structure, such as amino, pyridine, and pyrimidine [22,23]. This process usually involves a coordination bond between a Lewis base (the lone pair on the pyridine nitrogen) and a Lewis acid (the Pb(II) center), which passivates trap states.

Conventionally, SAMs are used as interlayers with charge-transport materials in p-i-n perovskite solar cell architectures, or even as ultrathin hole-transporting layers. In this configuration, SAMs primarily influence the buried interface of the perovskite film. However, due to the microcrystalline morphology of perovskite absorbers, interface modification is required not only at the top surface of the thin film but also at the grain boundaries within the absorber. Bulk doping allows to passivate undercoordinated traps through coordination and ionic bonds as it was reported for ionic liquids[24] and small organic molecules[25].

In this work, we present the impact of the synthesized molecule on the thermal stability of perovskite solar cells employing a $CsFAPbI_3$ absorber in p-i-n solar cells. We focused on the bulk integration of a pyridine-based SAM to investigate phase resilience under elevated temperature conditions of 85°C. It was found that doping with TPA-Py improves the morphology of the buried interface and suppresses the dynamics of phase segregation. Comprehensive investigation of the transport and photovoltaic performance of the devices revealed from passivation effects. We also identified the key role of the modified wetting angle of the perovskite solution upon the addition of the pyridine-based SAM. The obtained results were deeply analyzed and discussed.

## 2. Results and discussion

In this work, we used 4-(pyridin-4-yl)triphenylamine (**TPA-Py**) for bulk interface modification in p-i-n perovskite solar cells, as presented in **Fig.1** with corresponding device schematics and molecular structure. TPA-Py was synthesized via Suzuki coupling reaction between 4-pyridinylboronic acid and 4-bromo-*N,N*-diphenylaniline under standard conditions using tetrakis(triphenylphosphine)palladium(0) as a catalyst and sodium carbonate as a base. The synthesized TPA-Py is characterized by values of HOMO and LUMO energy levels equal to –5.47 and –2.34 eV, respectively, typical for derivatives of triphenylamine; high destruction (282˚C) and melting (150˚C) temperatures, exceeding the annealing and operating temperatures of the device. Details of synthesis, NMR spectra, thermogravimetric, differential scanning calorimetry analysis, UV-Vis absorption and cyclic voltammograms data presented in Electronic Supplementary Information (**ESI**) in the **Fig. S1–S5**. Chemical quantum calculations revealed that the HOMO and LUMO are localized on the spatially separated triphenylamine (electron donor) and pyridine (electron acceptor) moieties (see **Fig. S6, ESI**). This results in a relatively high observed dipole moment (3.97 D). Fabricated perovskite solar cells had a following architecture: soda-lime glass/In$_2$O$_3$:SnO$_2$ (ITO – anode, 220 nm)/NiO (hole transporting layer, 20 nm)/TPATC (self-assembled monolayer)/ CsFAPbI$_3$ absorber with TPA-Py bulk doping (0.07 wt%) (450 nm)/C$_{60}$ (Electron transport layer, 35 nm)/Bathocuproine (BCP, hole-blocking interlayer, 8 nm)/Copper (cathode, 100 nm). Complementary solubility of perovskite precursors and TPA-Py allowed to use SAM as solution additive. To simplify the sample identification, we used the following names: "Control" for the reference perovskite film configuration and "Target" for the absorbers modified with TPA-Py.

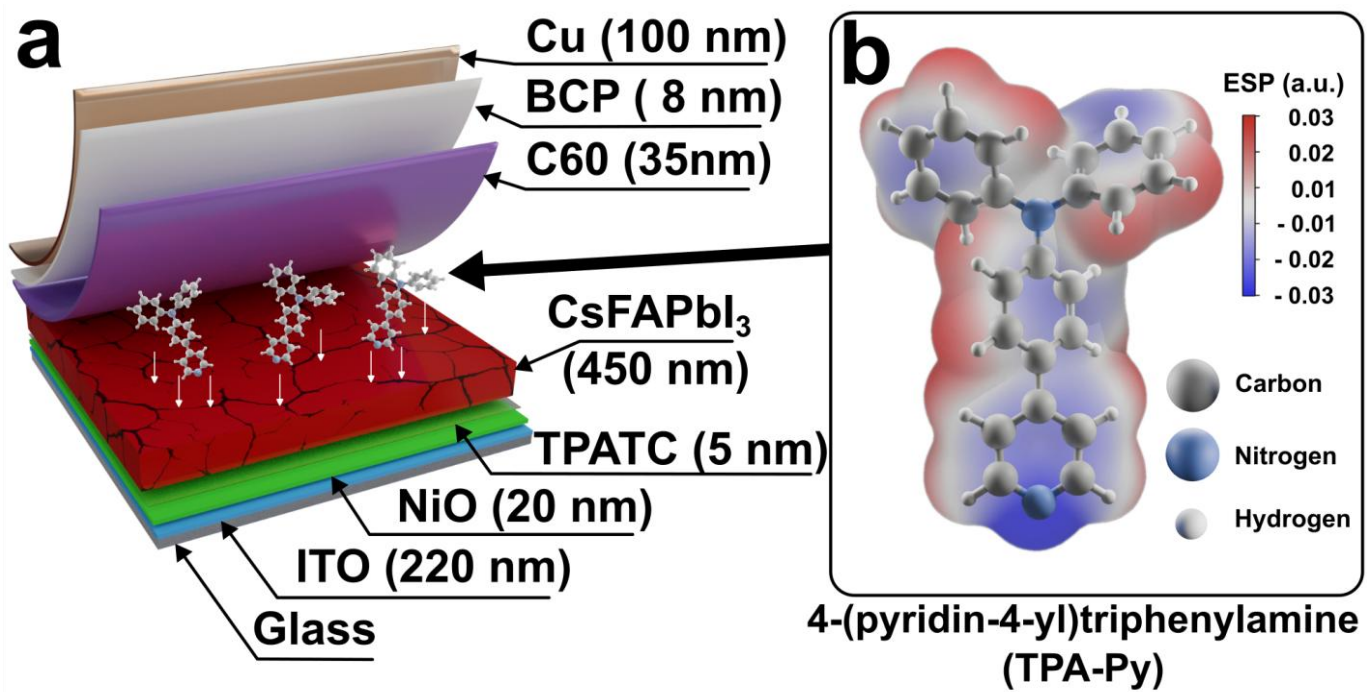

Figure 1 – Device schematics with bulk modification of perovskite absorber with TPA-Py (a), chemical structure and electrostatic potential surface of TPA-Py (b)

Structural properties of the absorber films of Control and Target were estimated with X-Ray diffraction (**XRD**). The corresponding diffractograms presented in the **fig.2(a).** Qualitative analysis of the data showed that both control and target samples predominantly composed of β-CsFAPbI$_3$ phase[26] with distinguish features at 13.95°, 19.81°, 24.34°, 26.35°, 28.16° and 31.59°. TPA-Py did not affect perovskite crystal lattice or inflict to strain effects. Lattice parameters: a = 8.93 Å and c = 6.33 Å equal for target and control samples. In Control sample small trace of hexagonal lead iodide can be found with peak at 12.62°, while Target showed the absence of the non-stoichiometric inclusions. To estimate the impact of TPA-Py bulk doping on the optical properties of perovskite films, we measured absorption and photoluminescence (PL) spectra (**Fig. S7**, **ESI**). The band gap values extracted from the Tauc plot[27] were 1.57 eV for both film configurations. The PL measurements showed a single peak at 780 nm of the same intensity for both sample configurations.

Surface morphology was investigated using atomic force microscopy (**AFM, fig. S8, ESI**). For the top surface of the crystallized absorbers, a minor increase in the average grain size was observed, from 160 nm for the Control sample to 170 nm for the Target one. A more notable change was found at the buried interface (in contact with the NiO/SAM interlayers). The control samples exhibited the formation of multiple isolated clusters with lateral dimensions of approximately 100 nm and heights up to 50 nm. In contrast, the Target samples reduced density of such defects with height not exceeding ~25 nm. We assume that the observed inhomogeneities could be related to the PbI$_2$, which was identified with XRD analysis.

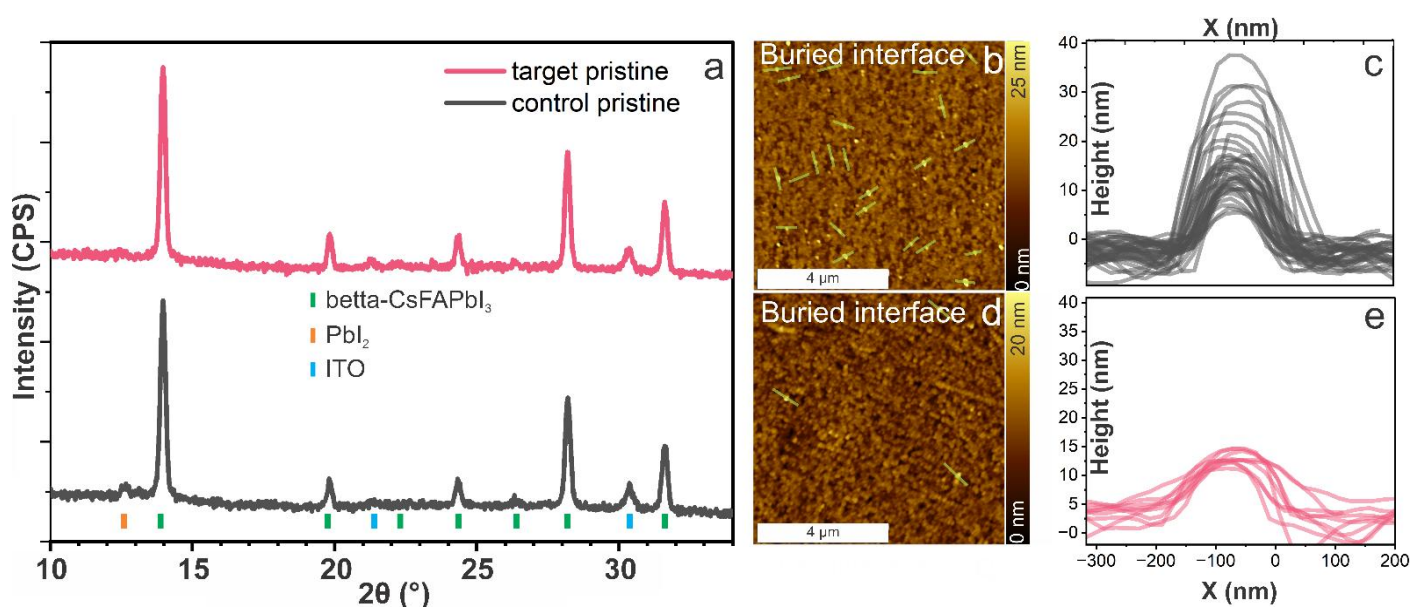

Figure 2 – XRD of control and target samples (a), buried interface (b) and defects profiles (c) for control sample and (d, e) for target sample

The changes in the crystal properties and/or doping could impact on the behavior of the trap states in the microcrystalline absorbers and shift the work-function (**W$_f$**) value. We used Kelvin probe force microscope (KPFM) method for measurement of the surface potential (**fig. S9**). TPA-Py modification induced the Fermi level pinning (**FLP**) effect[28–30] with reduction of ~0.44 eV compared to the Control sample, which revealed the rearrangement of the energy level alignment. SAMs typically possess intrinsic dipole moments[31]. The distribution of TPA-Py at the grain boundaries could alter the vacuum level and induces interface potential drop. The observed reduction in surface potential was sufficiently pronounced to suggest passivation of lead-related states, which typically exhibit energies close to the mid-gap region[32,33].

Nevertheless, the Fermi level became pinned near this interfacial state rather than shifting with the bulk composition[34]. The use of TPA-Py as a solution additive may allow pyridine to form adducts with $PbI_2$ fragments (involving the coordination of protonated pyridine with Pb-I complexes)[35]. Consequently, the emergence of charge-enriched/charge-depleted regions at inter-grain boundaries can induce local band bending [36,37].

The effect of SAM incorporation into the bulk of the microcrystalline absorber on charge-carrier transport and the dynamics of corrosion processes was evaluated through measurements of the output parameters of solar cells exposed to thermal stress to induce degradation. Analysis of the photoelectric performance for perovskite solar cells was made under standard conditions of illumination (AM 1.5 G spectra, 25°C and irradiance of 100 mW/cm$^2$ provided by Xe solar simulator of AAA class, see ESI for the details). The box-charts with statistical distribution of the voltampere characteristics (open circuit voltage - $V_{oc}$; short circuit current density - $J_{sc}$; filling factor – FF and power conversion efficiency (PCE) presented in the **fig.S10 (ESI)**. Measurements were performed under ambient conditions for the encapsulated devices. Controls PSCs had an average (maximum) $V_{oc}$ of 1.106 V (1.136 V) maximum , while for Target samples the value increased to 1.125 V (1.139 V). Average $J_{sc}$ were almost equal for both device configurations ~22.9 mA/cm$^2$. The measured spectra of external quantum efficiency supported the observed data (integrated value of $J_{sc}$ was ~22 mA/cm$^2$ , **Fig.S11** in **ESI**). We observed a minor improvement of FF for TPA-Py modified devices with raise from 76.5 to 77.5%. As result, the average PCE for Target improved from 19.4 to 20.1%. Both device configurations showed a slight hysteresis in JV curves (**Fig.S12** in **ESI**).Champion devices exhibited relevant values of 20.6 and 21.3%, respectively (JV curves presented in the **fig.3(a)**). The key-effect observed with introducing TPA-Py into the bulk of the solar cell absorber was an increase in the $V_{oc}$, which is governed by the position of the quasi-Fermi levels under illumination and the recombination dynamics. So, the induced local bend-bending at CsFAPbI$_3$/TPA-Py interface improved the charge-carrier splitting and possibly suppressed the trapping. The stabilization of the output performance over 10 minutes was estimated via maximum power point tracking measurements under standard illumination conditions (**Fig. S13**, **ESI**). For both device configurations, the power output ($P_{max}$) and $J_{sc}$ showed minimal changes over time. The amplitude of these changes was no more than 1.5% for $P_{max}$ and within 4% for $J_{sc}$.

The main focus of the current study is directed at the performance of devices under continuous thermal stress. We followed to the conditions of ISOS-D-2 protocol[38] with heating of the samples up to 85 ± 1.5°C in ambient atmosphere (Relative humidity was in the 30 – 60% range). The stability of the device performance) was estimated throughout the $T_{80}$ period (a period corresponding to a 20% relative loss of the device's maximum power output). Trend analysis of JV parameters (**Fig. 3b-d**) showed that $J_{sc}$ has a major impact on the evolution of PCE under thermal stress. Both device configurations have stable $J_{sc}$ values up to

200h of the test. A sharp drop was observed for Control configuration at 200-240h range with corresponding $T_{80}$ for $J_{sc}$ at 218h. Notably, after rapid decrease, the $J_{sc}$ showed a plateau of stable values at ~70-75% relatively to initial parameters. The Target device showed a smoother trend of $J_{sc}$. Finally, the PCE parameters were calculated as 260h for Control and 560h for Target samples. Comparison of the obtained results with similar studies in the field of thermal stabilization shows that bulk modification with SAMs is on par with conventional HTL surface engineering[39] or vapor-based treatments[5].

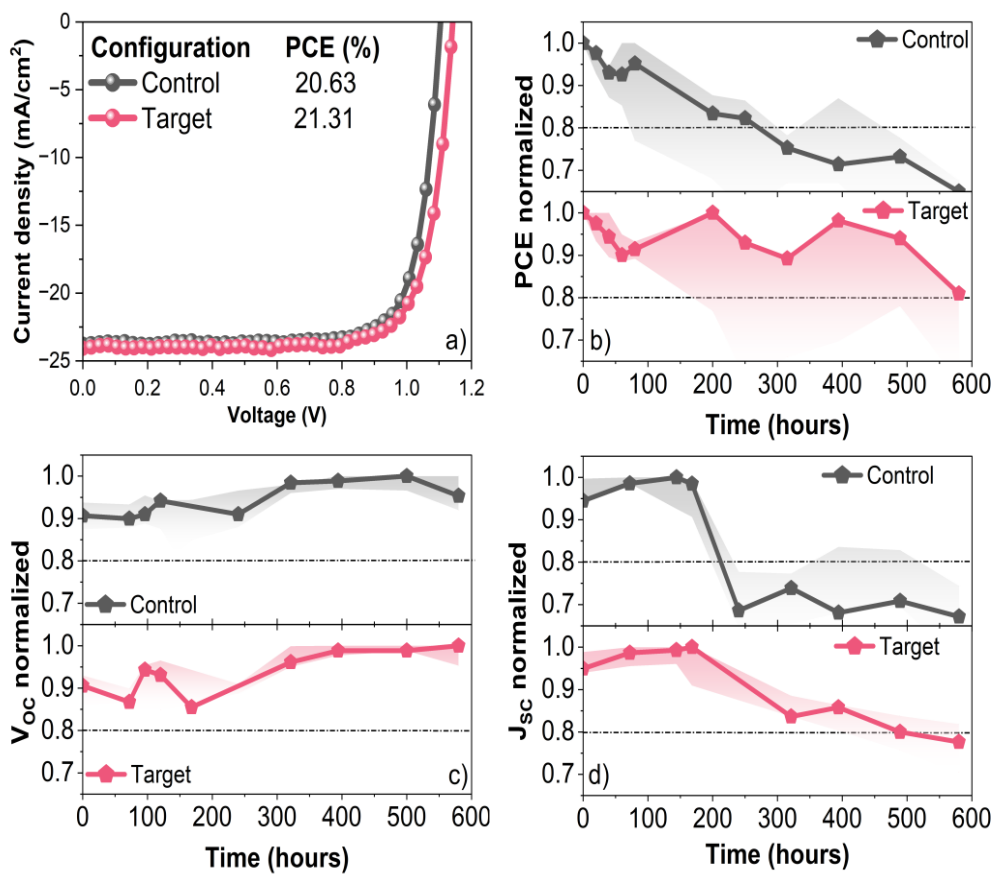

Figure 3 – The champion JV curves for the fabricated PSCs (a), PCE (b), $V_{oc}$ (c) and $J_{sc}$(d) stability trends for the devices under thermal stress under 85°C. The curves on the figures (b)-(c) represents the performance of the device with the best PCE, while semitransparent background shows the statistical data distribution

During thermal stress testing, additional photo-induced voltage transients (PIVTs) and admittance (AS) measurements were performed to probe trap behavior over a wide temperature range and varying heat exposure durations. PIVTs method aims to monitor $V_{OC}$ relaxation kinetics by tracking the device response to light pulses at different temperatures. Upon illumination, the solar cell reaches its steady-state $V_{OC}$. When the light is switched off, non-equilibrium carriers typically recombine within nanoseconds, causing rapid $V_{OC}$ relaxation. In perovskite solar cells, this relaxation is slowed by mobile ions that contribute to ionic currents and screen electric fields. Tracking the time dependence of $V_{OC}$ after a light pulse thus provides insights into ion kinetics and enables estimation of $V_{OC}$ within defined temperature intervals. Representative PIVTs results are shown in **Fig. 4(a) (b)**, with admittance spectroscopy data in **Fig. 4 (c)**. In this study samples were illuminated with a 470 nm LED of approximately 250 mW/cm² optical power density. After the illumination pulse, the subsequent $V_{OC}$ relaxation was recorded. Measurements were performed over a temperature range of 200 – 340 K to make it possible to distinguish the contribution of mobile ions kinetics and to determine temperature dependent diffusion coefficient. Arrhenius plot and estimation of diffusion coefficient of mobile ions presented in the **Fig. 4(c)**. If the screening of the built-in electric field of the sample is governed by the presence of some mobile ion specie, then the characteristic time required for an ion to travel along Debye screening length $L_D$ is approximated by $\tau = \frac{L_D^2}{D}$, where $D$ is the diffusion coefficient. By correlating temperature dependent $V_{OC}$ kinetics with the slow kinetics of mobile ions the diffusion coefficient can be determined in PIVTS as well as direct ion induced capacitance measurements at frequency $\tau = \omega^{-1}$ [40,41].

$$\omega^{-1} = \tau \approx L_D^2/D = \frac{\epsilon\epsilon_0 k_B T}{q^2 N_i D_0} \exp\left(\frac{E_A}{k_B T}\right) \quad \rightarrow \quad D(T) \approx \frac{\epsilon\epsilon_0 k_B T}{q^2 N_i \tau} \quad (1)$$

where $\epsilon\epsilon_0$ – is material dielectric permittivity, $k_B$ – is Boltzmann constant, and $N_i$ – mobile ion concentration.

Both samples demonstrated mobile Ion 2 feature at 275-325 K temperature band with diffusion activation energy to be $E_a$=0.49 eV and $D_0$=1.2×10⁶ cm²/s (or $D_{300K}$=7.0×10⁻³ cm²/s) which also was found in the work [42]. Such high diffusivity values can be attributed to light-induced ion migration, as reported in [43–45] and could be related to the migration of FA (vacancy $V_{FA}$[46]). Target device exhibited additional Ion 1 feature at 200-250 K band with an activation energy to be $E_a$=0.45 eV and $D_0$=4.1 cm²/s (or $D_{300K}$=1.1×10⁻⁷ cm²/s) [47,48]. The amplitudes of the PIVTs signal for Ion 2 in the Control samples demonstrate a steep increase, reaching over 30% at 80 h of thermal stress exposure, whereas the signals for the two ions in the Target samples show a moderate rise over the thermal stress duration of 20-80 h. Ion 3 feature detected in all samples while performing Admittance Spectroscopy (see **Fig. 4(c)**, with parameters $E_a$=0.53 eV and $D_0$=10.9 cm²/s (or $D_{300K}$=1.4×10⁻⁸ cm²/s), typically reported in literature($V_I$[49],[50]])

In parallel, we evaluated the recombination lifetime of excess carriers via transient photovoltage measurements (TPV), as shown in fig**. 4(d).** We analyzed the fall mode measurements to gather insights into lifetime values and effects of enhanced recombination after thermal treatment. The data capture falling profiles under open-circuit conditions, fall ($t_f$) times were defined as the interval between 90% and 10% of the saturated signal. At the start of the stability test, the Control device exhibited $t_f$ = 6.3 μs, while the Target showed a slightly higher value of 7.6 μs, indicating longer carrier lifetimes. Both devices displayed decreasing $t_f$ with thermal stress duration. Statistical analysis confirmed a modest reduction for the Control, where $t_f$ dropped to 3.4 μs after 300 h, compared to 4.4 μs for the Target. Representative fall mode decay plots presented in the **Fig.S14 (ESI)**. The observed trends indicate slower recombination in the Target device, consistent with the higher quasi-Fermi level splitting ($V_{OC}$ gain) seen in AM 1.5 G JV characteristics[42].

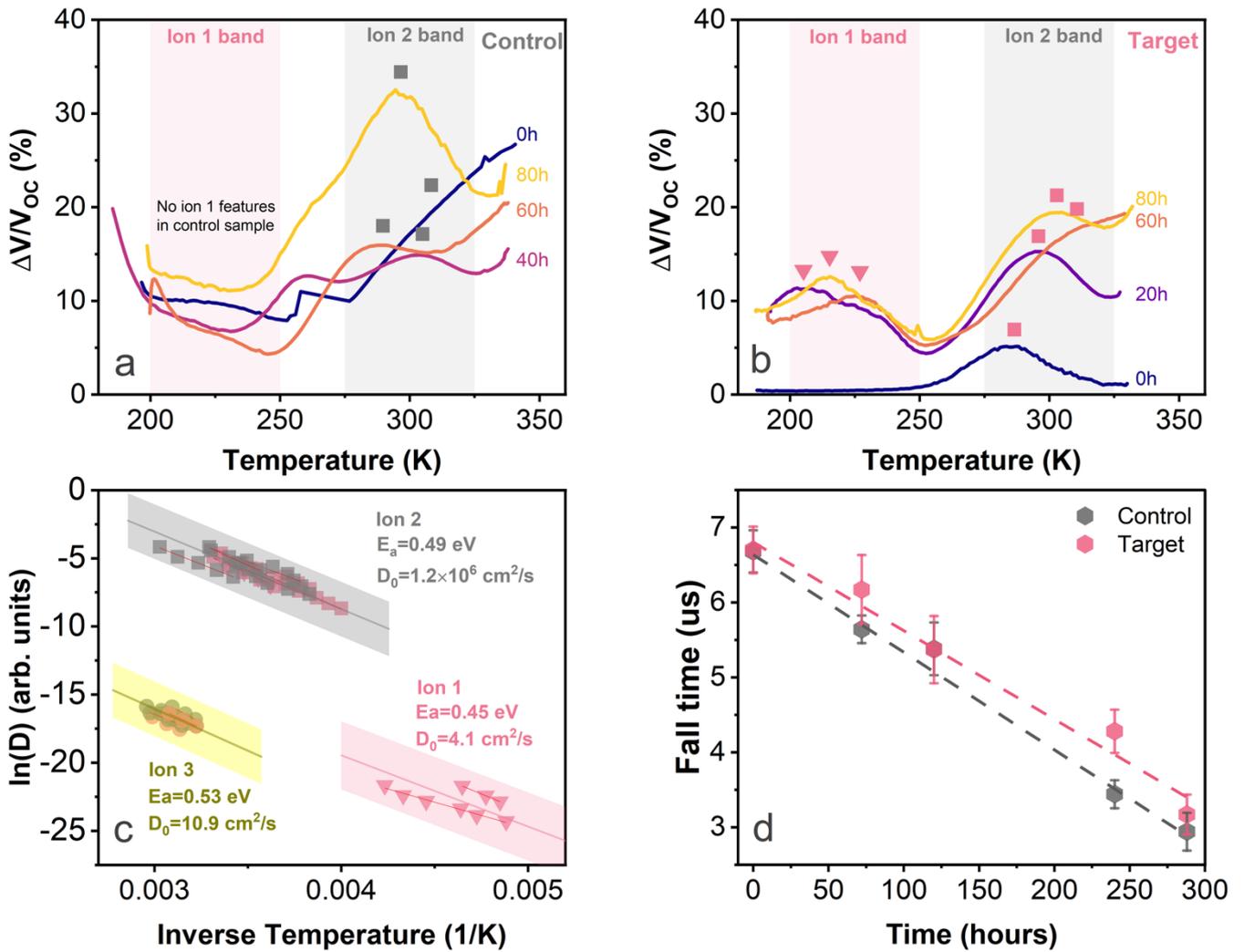

**Figure 4** – $\Delta V/V_{OC}$ relaxation amplitudes measured as a function of temperature (PIVTS method) for (a) control and (b) target samples after different stress periods (c) Arrhenius plot and estimation of diffusion coefficient of mobile ions. (d) Dependence of fall time values extracted from TPV measurements for PSCs exposed to thermal stress

The crystallization of perovskite absorbers on nanocrystalline oxide surfaces strongly depends on the surface state, including parameters such as contact angle, wettability, and chemical activity[52]. In the context of the p-i-n device architecture employed in this work, it is important to emphasize that non-stoichiometric NiO and local inhomogeneities in the hole-transport SAM can induce chemical interactions with the organic cation of the perovskite lattice[53], leading to subsequent decomposition ($PbI_2$ formation)[54] and segregation (for instance, $CsPbI_3$ formation)[55]. Thermal stress serves as a key trigger for this corrosive process. To assess the effect of TPA-Py on thermal stability of absorber films, SEM (scanning electron microscopy) and EDX (energy-dispersive X-ray spectroscopy) analyses were performed on Control and Target devices during 600 h of heat stress under encapsulation. To gather the data, we made the delamination of the devices from the batch (ETL, HBL with cathode were peeled off) and studied the surface of the perovskite films. **Fig. 5(a,b)** presents SEM micrographs of the control and modified samples. Both surfaces exhibit bright defect features distributed across the pixel area. Quantitative analysis indicates that defect clusters cover 19.3% of the active surface in the control device, while in the TPA-Py-treated device the coverage is limited to 8.4%. EDX analysis (**Fig. 5(c)**) demonstrates an enrichment of Cs within the defect clusters accompanied by Cs depletion in the surrounding perovskite matrix. Micrometer-sized clusters were also visible under optical microscopy (**Fig. S15** in **ESI**). The area fraction of these features increases progressively with prolonged thermal exposure (Figure 5(d)). However, the growth rate was suppressed in the TPA-Py-modified sample. Notably, that rapid growth of the cluster fraction was observed in the period 150-200h of thermal stress, which follows the trend of $J_{sc}$ drop in the device output performance (see **fig. 3(d)**).

Further structural characterization of the peeled films using X-ray diffraction (XRD, **Fig. 5e,f**) confirms that the β-CsFAPbI$_3$ phase remains predominant after stress, although a pronounced increase in PbI$_2$ content is detected, consistent with thermally induced decomposition. This degradation pathway is attributed to the volatilization of the organic FA$^+$ cation, leading to irreversible perovskite phase breakdown. Additionally, diffraction patterns reveal traces of δ-CsPbI$_3$ in stressed samples. Correlating these results with the EDX data, it can be inferred that TPA-Py mitigates Cs-rich phase segregation and thereby contributes to enhanced thermal stability by suppressing FA$^+$. It has been demonstrated that, due to the presence of a pronounced dipole moment, small molecules such as TPA-Py are capable of passivating uncompensated ionic states[36,37]. Furthermore, given that pyridine in TPA-Py is a Lewis acid, the molecule is capable of various coordination interactions with FAI[56]. These interactions were demonstrated using $^1$H NMR spectroscopy method (see **ESI**, **Fig. S16**).

The observed data also correlates with PIVTS measurements. The addition of TPA-Py at the grain boundaries of the absorber can be directly related to presence of the Ion 1 feature (0.45 eV) in Target samples (absence in Control samples), which originates from accumulation of mobile ion species trapped inside grains and results in effective built-in field screening along the grains[57,58]. Trapping mobile ions inside grains favors field screening and also limits their migration across device active area, which reduces the supply of phase clusters formation and helps maintain cell efficiency (Fig.5(d)). The emerging amplitude of the PIVTS signal for Ion 2 after 80 h of thermal stress is 1.5 times higher in the Control sample, highlighting the ability of TPA-Py to effectively 'shut' the grain boundaries limiting ion migration and also preventing formation of additional phase clusters[59].

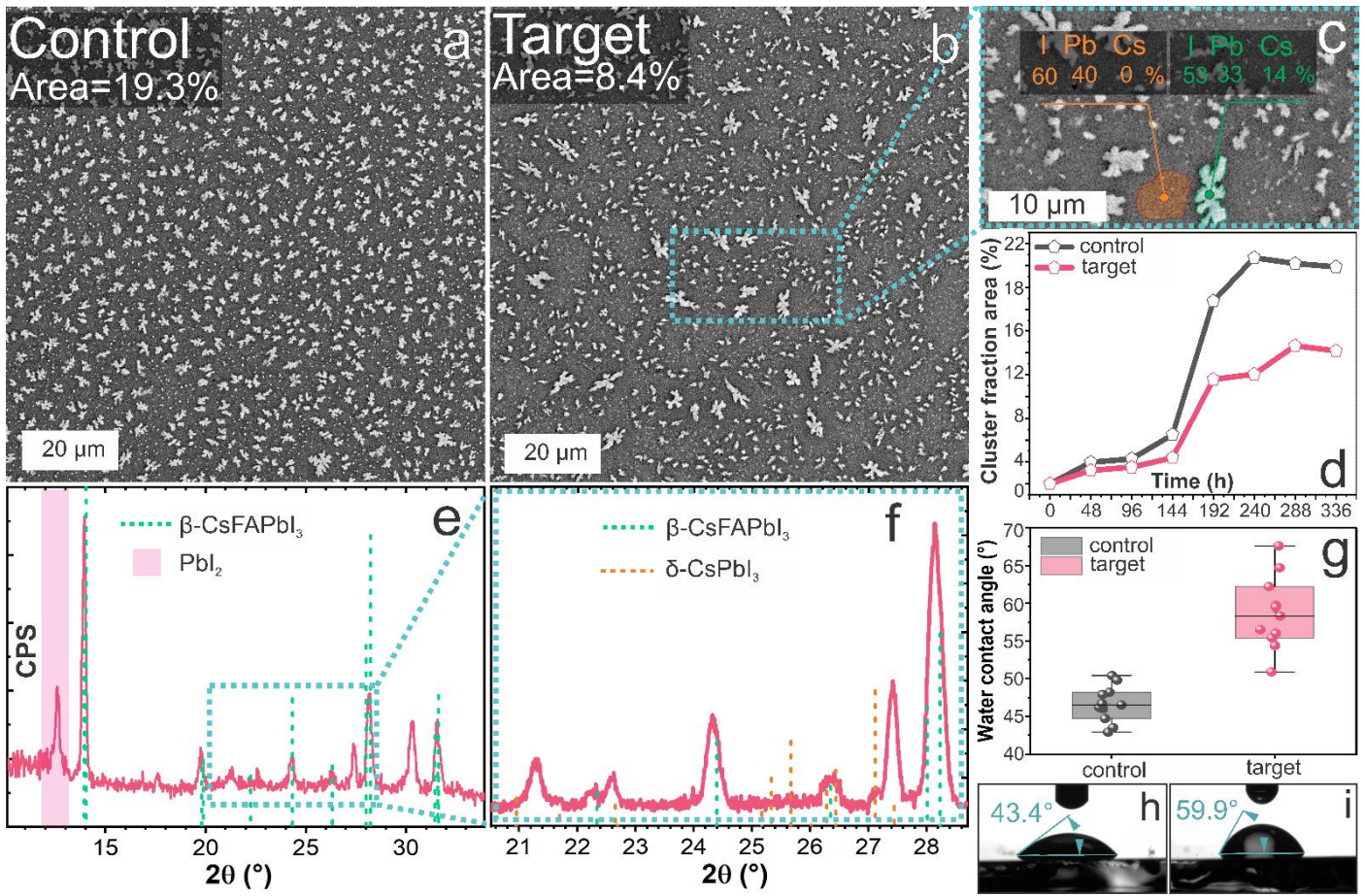

Figure 5 – SEM images of control (a) and target (b) pixel area after 600+ h of thermal stress 85˚C, (c) EDX analysis of clusters, evolution of clusters area fraction in control and target devices (d), XRD analysis of perovskite film after thermal stress (e, f), water contact angle on perovskite surface(g), images of water droplets on control (h) and target(i) films

Passivation of perovskite interfaces with hydrophobic groups was expected to enhance the overall hydrophobicity of the perovskite surface. To verify the passivation effect, water contact angle measurements were performed on perovskite film samples (Figure 5g). In Figure 5h,i representative water droplet profiles on the surface of perovskite films with and without the incorporation of TPA-Py are shown. TPA-Py in the perovskite film increased the average water contact angle from 46˚ to 58˚, indicating a significant enhancement of surface hydrophobicity and confirming the effectiveness of passivation by the organic molecule. Improved hydrophobicity contributes to higher chemical stability of the perovskite when exposed to ambient atmospheric conditions[60–62].

Typical approaches employing SAMs are primarily aimed at forming an interlayer between HTL and the absorber. Achieving uniform monolayers via solution processing remains a non-trivial technical challenge. Conventional use of SAMs often results in pinholes and/or clusters with high hydrophobicity when scaling up the active area of devices[52]. Perovskite decomposition and phase segregation, accelerated by thermal stress, arise from direct contact with nanocrystalline oxides within the device structure, such as ITO and NiO. The introduction of SAM interlayers at the hole-transport interface minimizes the exposure of the absorber to nanocrystalline oxides of the substrate. On the other hand, this strategy has limited impact on the grain boundaries of the perovskite film, both at the surface and within the bulk. Incorporating SAMs directly into the precursor solution allows direct passivation of potential sites for corrosive and non-radiative processes. However, harsh thermal stress may also induce disruption of SAMs, which necessitates the formation of robust chemical bonds with the perovskite lattice and careful molecular design to suppress thermal motion effects within the corresponding interlayers.

Thermal decomposition of CsFAbI$_3$ into CsPbI$_3$ and PbI$_2$ is a defect-mediated process that is exacerbated at the buried interface due to chemical incompatibility, lattice strain, and interfacial reactions. The formation of strong dipoles with TPA-Py SAM to screen ion migration and phase resilience clearly demonstrated the importance of the interface within the bulk of the perovskite film. Further progress requires increasing the thermal stability threshold to 140-150 °C, which is characteristic of the standard lamination processes used in solar panel fabrication. The presented strategy can also be combined with the fabrication of ultrathin heterostructures incorporating layered perovskites[63,64], which provide effective, though not complete, suppression of ionic migration.

## 3. Conclusions

In this work, we presented important insights into bulk modification of double-cation perovskite absorbers using SAMs. The structure of the asymmetric molecule included a triphenylamine electron-donor core and a pyridine electron-acceptor moiety. Such a molecular design allows specific interactions between TPA-Py and perovskite and enables the molecule to have a high dipole moment, resulting in inter-grain band bending. The compatibility of solution-processed TPA-Py with the perovskite absorber improved the buried interface with the HTL, where the formation of PbI$_2$ defect clusters was significantly reduced. Modification of surface energetics led to a reconfiguration of defect states, while the use of TPA-Py enabled the screening of ionic defect migration with an activation energy of 0.45 eV, as determined by PIVTS measurements. Consequently, the comprehensive impact of TPA-Py on interfacial defects within the CsFAPbI$_3$ absorber resulted in suppressed non-radiative recombination dynamics, an increase in open-circuit voltage to 1.14 V, and a power conversion efficiency of 21.3%. The key feature of employing bulk TPA-Py SAM modification, however, was the enhancement of device thermal stability under harsh heating conditions (ISOS-D-2, 85 °C). Batch trend analysis revealed an increase in T80 lifetime from 260 h to over 560 h. These improvements in stability were attributed to enhanced phase resistance against CsPbI$_3$ formation, which primarily governed the photocurrent decay dynamics under thermal stress. The obtained results highlight the promise of bulk perovskite modification as an effective strategy to overcome one of the most critical bottlenecks-intense corrosion processes under heating and also demonstrating strong potential for integration into scalable fabrication routes such as slot-die and inkjet printing.

**CRediT authorship contribution statement**

**Ekaterina A. Ilicheva:** Writing – original draft, Visualization, Validation, Methodology, Investigation, Formal analysis, Data curation, Conceptualization; **Irina A. Chuyko:** Resources, Writing – original draft, Validation, Investigation, Formal analysis; **Lev O. Luchnikov:** Writing – original draft, Visualization, Validation, Methodology, Investigation, Formal analysis, Data curation; **Polina K. Sukhorukova:** Writing – original draft, Investigation, Formal analysis, Data curation; **Nikita S. Saratovsky:** Investigation, Formal analysis; **Anton A. Vasilev:** Investigation, Formal analysis, Data curation; **Luiza Alexanyan:** Formal analysis, Data curation; **Anna A. Zarudnyaya:** Formal analysis, Data curation; **Dmitri Yur. Dorofeev:** Formal analysis, Data curation; **Sergey S. Kozlov:** Formal analysis, Data curation; **Andrey P. Morozov:** Formal analysis, Data curation; **Dmitry S. Muratov:** Visualization, Data curation; **Yuriy N. Luponosov:** Writing – original draft, Formal analysis, Data curation, Supervision, Conceptualization; **Danila S. Saranin:** Writing – original draft, Visualization, Validation, Supervision, Resources, Project administration, Investigation, Funding acquisition, Formal analysis, Data curation, Conceptualization.

**Declaration of competing interest**

The authors declare no competing financial interest.

**Supporting Information**

The electronic supplementary material contains the following data:

**Acknowledgements**

The authors gratefully acknowledge the financial support from the Russian Science Foundation (RSF) with grant № 22-19-00812-P. Authors acknowledge Michael D. Khitrov (ISPM RAS) for carrying out DFT calculations.

# Supporting Information

**Enhanced thermal stability of inverted perovskite solar cells by bulky passivation with pyridine-functionalized triphenylamine**

*Ekaterina A. Ilicheva[1,§], Irina A. Chuyko[2,§], Lev O. Luchnikov[1], Polina K. Sukhorukova[1,2], Nikita S. Saratovsky[2], Anton A. Vasilev[3], Luiza Alexanyan[3], Anna A. Zarudnyaya[1], Dmitri Yur. Dorofeev[1], Sergey S. Kozlov[4], Andrey P. Morozov[1], Dmitry S. Muratov[5], Yuriy N. Luponosov[2,*], and Danila S. Saranin[1,*]*

[1]LASE – Laboratory of Advanced Solar Energy, NUST MISIS, 119049 Moscow, Russia

[2]Enikolopov Institute of Synthetic Polymeric Materials of the Russian Academy of Sciences (ISPM RAS), Profsoyuznaya St. 70, Moscow, 117393, Russia

[3]Department of semiconductor electronics and device physics, NUST MISIS, 119049 Moscow, Russia

[4]Laboratory of Solar Photoconverters, Emanuel Institute of Biochemical Physics, Russian Academy of Sciences, 119334 Moscow, Russia

[5]Department of Chemistry, University of Turin, 10125, Turin, Italy

[§]The authors contributed equally to this work

**Corresponding authors:**

Dr. Yu. N. Luponosov luponosov@ispm.ru,

Dr. Danila S. Saranin saranin.ds@misis.ru.

**Experimental details:**

*Materials for synthesis*

4-Pyridinylboronic acid (Macklin), palladium-tetrakis(triphenylphosphine) (Pd(PPh$_3$)$_4$, LEAPChem), sodium carbonate. 4-Bromo-*N,N*-diphenylaniline was prepared as described elsewhere Ref. [S1]. Toluene, ethanol and other solvents were purified and dried according to known methods.

*Materials for PSCs fabrication*

Devices were fabricated on In$_2$O$_3$: SnO$_2$ (ITO) coated glass (R$_{sheet}$ < 10 Ohm/sq) from Zhuhai Kaivo company (China). NiCl2·6H2O (from ReaktivTorg 99+% purity) used for hole-transport material fabrication. Formamidinium iodide (FAI, 99.9%) was purchased from LLC Polysense (Russia). TPATC (99+% purity, synthesized with respect to our previous work [S2] at the ISPM RAS (Russia)) used for SAM fabrication on interface NiO/Perovskite. Lead Iodide (99.9 %), Cesium iodide (99.99 %), trace metals basis from LLC Lanhit (Russia). C60 (99.8 %) was purchased from MST NANO (Russia). BCP (99.0%) was purchased from Leap Chem Co. (China). The organic solvents: dimethylformamide (DMF), N-methyl-2-pyrrolidone (NMP), chlorobenzene (CB), isopropyl alcohol (IPA) and 2-methoxyethanol (2-ME) were purchased in anhydrous from Sigma Aldrich and used as received without further purification. Ethyl acetate (EAC) was purchased from LLC ALDOSA (Russia).

*Synthetic procedures of TPA-Py*

Pd(PPh$_3$)$_4$ (0.160 g, 0.1 mmol) was added to a mixture of 4-bromo-*N,N*-diphenylaniline (1.5 g, 4.6 mmol) and 4-pyridineboronic acid (0.682 g, 5.6 mmol). Then 38 ml of toluene, 6 ml of ethanol and 8.3 ml of sodium carbonate were added. The reaction mixture was boiled in an inert atmosphere for 18 hours using microwave radiation. After completion of the reaction, 100 ml of toluene and 200 ml of water were added to the reaction mixture, the organic phase was separated and washed three times with water. After purification by column chromatography on silica gel (eluent - ethyl acetate: petroleum ether = 2:5), followed by recrystallization from dichloromethane with ethanol, the product was obtained as a solid yellow substance (yield: 0.88 g, 59%). $^1$H NMR (250 MHz, CDCl$_3$, δ, ppm): 7.03–7.17 (overlapping peaks, 8H), 7.25–7.35 (overlapping peaks, 4H), 7.43–7.55 (overlapping peaks, 4H), 8.60 (dd, 2H, $J_1$ = 1.83 Hz, $J_2$ = 6.10 Hz). $^{13}$C NMR (75 MHz, CDCl$_3$, δ, ppm): 120.85, 122.90, 123.59, 124.97, 127.63, 129.42, 130.92, 147.23, 147.61, 148.93, 150.19. MALDI-TOF MS: found m/z 322.06; calcd. for [M]$^+$ 322.15.

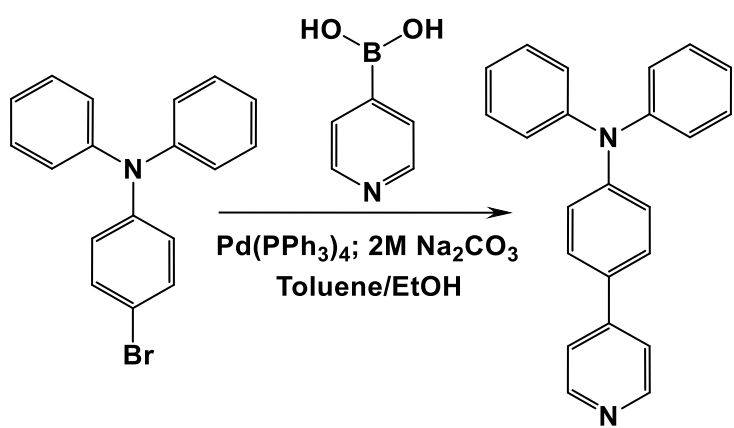

**Figure S1** – Synthesis of **TPA-Py**.

*Preparation of the Precursors*

Nickel oxide precursor ink was prepared by dissolving the NiCl2·6H2O powder in 2-ME with a concentration of 50 mg mL$^{-1}$. Then, the green solution was heated at 75 °C in 2 h after adding nitric acid with a concentration of 25 μL mL$^{-1}$ relative to 2-ME. The TPATC solution was prepared by dissolving 1 mg of TPATC in 1 mL of CB, and then the solution was heated at 50 °C in 1 h before using. To fabricate the perovskite ink Cs$_{0.2}$FA$_{0.8}$PbI$_{2.93}$Cl$_{0.07}$ (text abbreviation CsFAPbI$_3$) we used the halide salts (CsCl, CsI, FAI, PbI2) with molar proportions 0.07:0.13:0.2:1. The resulting mixture was dissolved in a DMF:NMP (volume ratio 640:360) with a concentration of 1.35 M and stirred at a temperature of 50 °C for 1 h. Before coating the precursor, the solution was filtered with 0.45 μm PTFE filters. For target films, after complete dissolution, in the perovskite solution TPA-Py was added in the concentration of 0.5 mg/ml and stirred for 30 min before coating.

*Device fabrication*

The p-i-n-structured PSCs were fabricated with the following stacks ITO/NiO/TPATC/CsFAPbI3/C60/BCP/Cu (control) and ITO/NiO/TPATC/CsFAPbI3:TPA-Py/C60/BCP/Cu (target). The ITO substrates were cut in a dimension of 25 × 25 cm$^2$ and patterned by UV–laser to isolate 5 mm wide semitransparent ITO electrodes. Firstly, the ITO substrates were cleaned with detergent, de-ionized water, acetone, and IPA in an ultrasonic bath. Then the substrates were activated under UV-ozone irradiation for 30 min. NiCl2·6H2O precursor for NiO HTM film was spin-coated at 4000 RPMs (30 s), dried at 120 °C (10 min), and annealed at 300 °C (1 h) in the ambient atmosphere with a relative humidity not exceeding 40%. The deposition and crystallization processes of perovskite layers were conducted inside glove boxes with an inert nitrogen atmosphere. CsFAPbI3 or CsFAPbI3:TPA-Py film was crystallized on top of HTL with a solvent-engineering method. The perovskite precursor was spin-coated with the following ramp: (5s–3000 rpm, 22 s–5000 rpm).

420 μL of EAC were dropped on the substrate on the 12th second after starting of the second rotation step. Then the substrates were annealed at 85 °C (1 min) and 105 °C (30 min) to form the appropriate perovskite phase. C60 and BCP was deposited with the thermal evaporation method at $10^{-6}$ Torr vacuum level. The copper was also deposited with thermal evaporation through a shadow mask to form a 3 mm wide copper electrode. The single cells were fabricated with 0.15 cm$^2$ active area measured equal to the intersection area of ITO and top Cu electrodes. After evaporation of Cu back electrode, we removed the NiO/Perovskite/C60/BCP stack from the edges of the glass substrate, using a razor blade. The safety area from the substrate edge to p-i-n device stacks was approximately 5.5-6 mm from each side. Then, we covered the pixels of PSCs with Kapton polyimide tape, ensuring full coverage with an additional 1 mm overlap on each side (15x15 mm). For the final encapsulation we used UV-curable epoxy (Ossila UV epoxy) and cover glass. UV epoxy was dispensed in small drops (approximately 100 μl per substrate) onto the polyimide tape, followed by the placement of a 17x17 mm cover glass. The cover glass was slightly larger than the tape to enclose the device structure. The glass was pressed to distribute the epoxy evenly underneath. The final encapsulation step involves curing the devices under UV-A lamp for 30 minutes, ensuring the epoxy is fully cured.

*Characterization*

**NMR spectra.** $^1$H NMR spectra were recorded using a "Bruker WP-250 SY" spectrometer, working at a frequency of 250 MHz and using CDCl$_3$ (7.25 ppm) or DMSO-d6 (2.50 ppm) signals as the internal standard. $^{13}$C NMR spectra were recorded using a "Bruker Avance II 300" spectrometer, working at a frequency of 75 MHz. In the case of $^1$H NMR spectroscopy, the compounds to be analyzed were taken in the form of 1% solutions in CDCl$_3$. In the case of $^{13}$C NMR spectroscopy, the compounds to be analyzed were taken in the form of 5% solutions in CDCl$_3$. The spectra were then processed on the computer using the "ACD Labs" software.

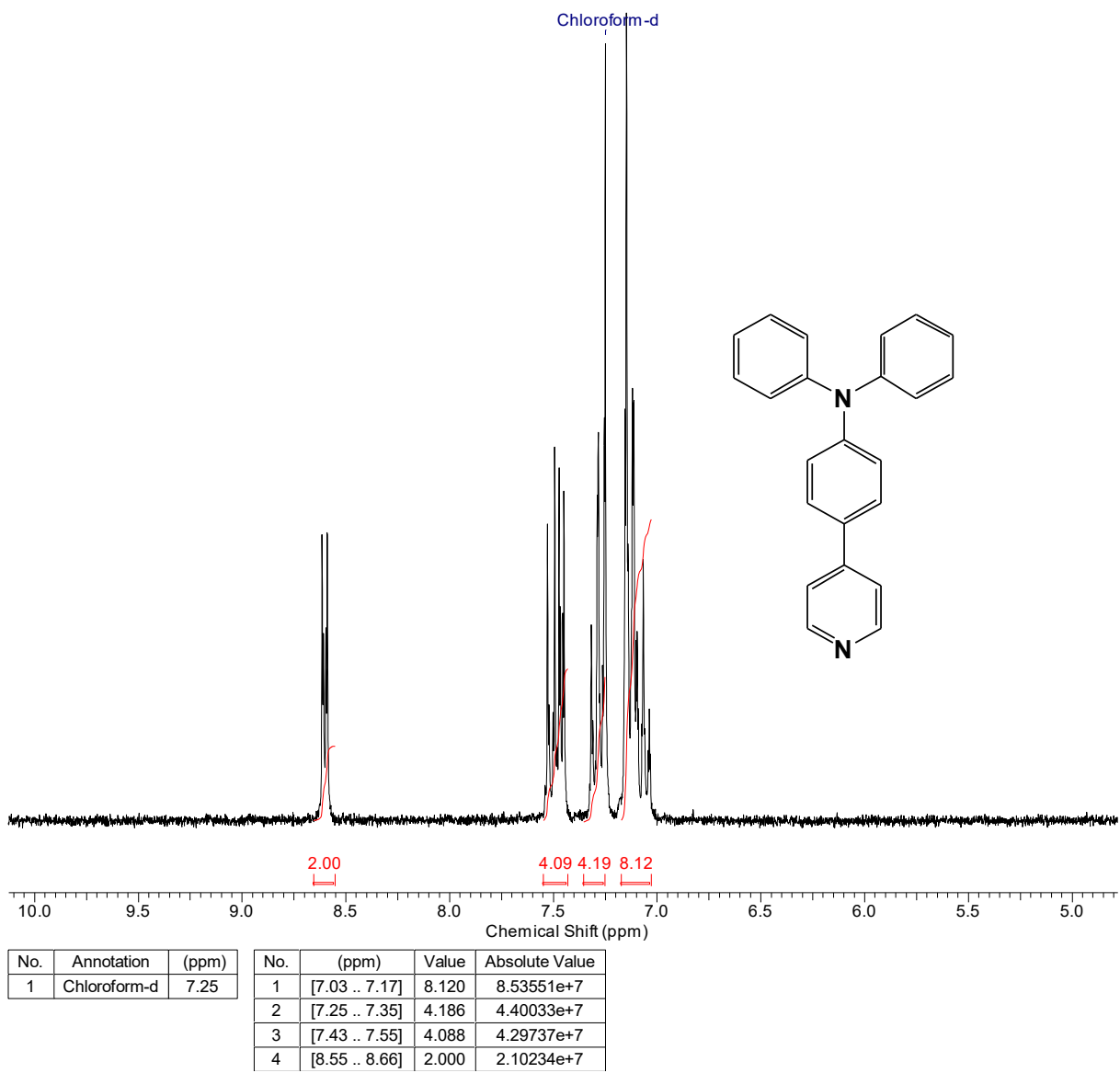

| No. | Annotation | (ppm) |
|---|---|---|
| 1 | Chloroform-d | 7.25 |

| No. | (ppm) | Value | Absolute Value |
|---|---|---|---|
| 1 | [7.03 .. 7.17] | 8.120 | 8.53551e+7 |
| 2 | [7.25 .. 7.35] | 4.186 | 4.40033e+7 |
| 3 | [7.43 .. 7.55] | 4.088 | 4.29737e+7 |
| 4 | [8.55 .. 8.66] | 2.000 | 2.10234e+7 |

**Figure S2**. $^1$H NMR spectrum of TPA-Py in CDCl$_3$.

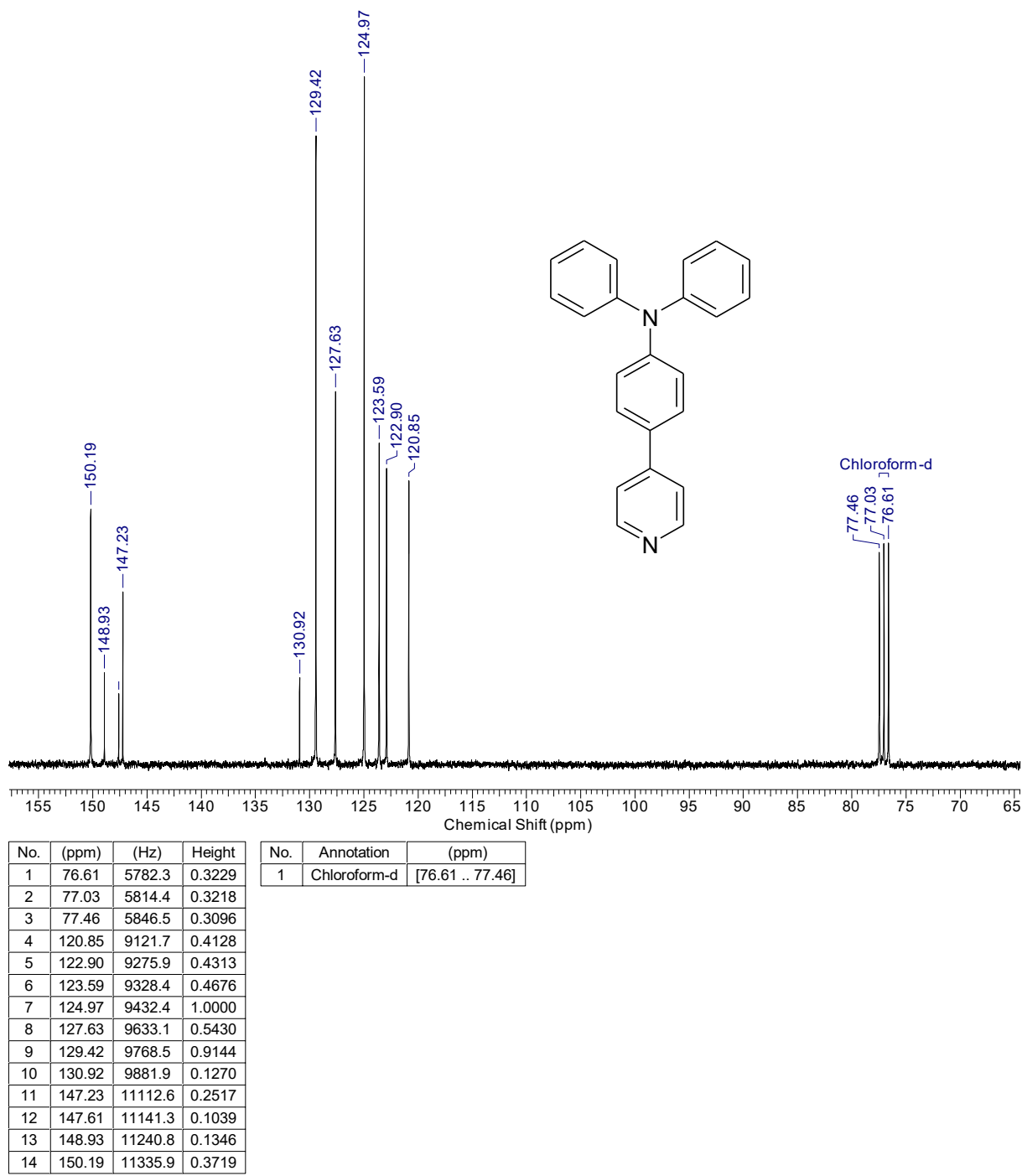

| No. | (ppm) | (Hz) | Height |
|---|---|---|---|
| 1 | 76.61 | 5782.3 | 0.3229 |
| 2 | 77.03 | 5814.4 | 0.3218 |
| 3 | 77.46 | 5846.5 | 0.3096 |
| 4 | 120.85 | 9121.7 | 0.4128 |
| 5 | 122.90 | 9275.9 | 0.4313 |
| 6 | 123.59 | 9328.4 | 0.4676 |
| 7 | 124.97 | 9432.4 | 1.0000 |
| 8 | 127.63 | 9633.1 | 0.5430 |
| 9 | 129.42 | 9768.5 | 0.9144 |
| 10 | 130.92 | 9881.9 | 0.1270 |
| 11 | 147.23 | 11112.6 | 0.2517 |
| 12 | 147.61 | 11141.3 | 0.1039 |
| 13 | 148.93 | 11240.8 | 0.1346 |
| 14 | 150.19 | 11335.9 | 0.3719 |

| No. | Annotation | (ppm) |
|---|---|---|
| 1 | Chloroform-d | [76.61 .. 77.46] |

**Figure S3**. $^{13}$C NMR spectrum of TPA-Py in CDCl$_3$.

**Mass-spectra.** Mass-spectra (MALDI-TOF) were registered on a "Autoflex II Bruker" (resolution FWHM 18000), equipped with a nitrogen laser (work wavelength 337 nm) and time-of-flight mass-detector working in the reflections mode. The accelerating voltage was 20 kV. Samples were applied to a polished stainless-steel substrate. Spectrum was recorded in the positive ion mode. The resulting spectrum was the sum of 300 spectra obtained at different points of the sample. 2,5-Dihydroxybenzoic acid (DHB) (Acros, 99%) and α-cyano-4-hydroxycinnamic acid (HCCA) (Acros, 99%) were used as matrices.

**TGA.** Thermogravimetric analysis (TGA) was carried out in dynamic mode in 30 ÷ 600°C interval using a "Mettler Toledo TG50" system equipped with M3 microbalance allowing measuring the weight of samples in 0–150 mg range with 1 μg precision. Heating/cooling rate was chosen to be 10 °C/min. Every compound was studied twice: in the air and an under argon flow of 200 mL/min.

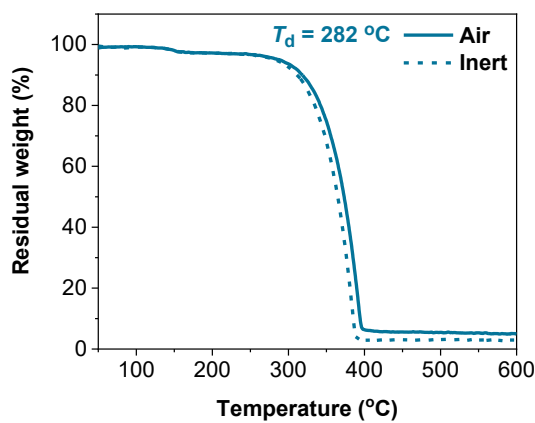

Figure S3 – TGA curves in air and inert atmosphere of **TPA-Py**

**DSC.** Differential scanning calorimetry (DCS) scans were obtained with a "Mettler Toledo DSC30" system with 10 °C/min heating/cooling rate in temperature range of +20–250 °C for all compounds. The $N_2$ flow of 50 mL/min was used.

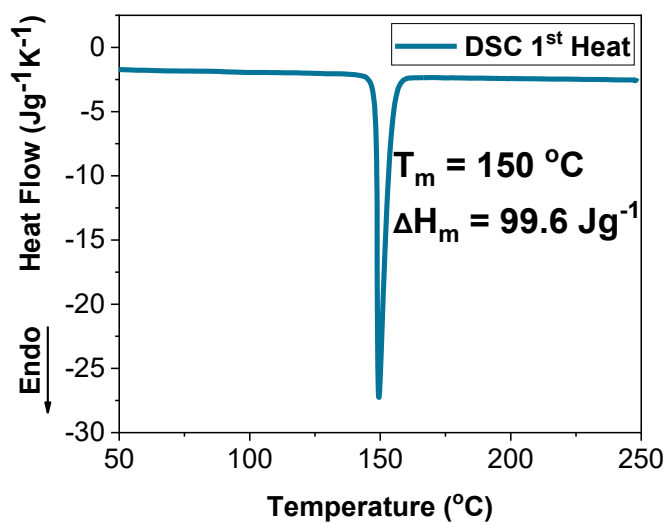

Figure S4 – The first DSC heating scans of **TPA-Py**

**UV-Vis steady state spectroscopy.** The absorption spectra of TPA-Py and blended solutions were recorded with a "Shimadzu UV-2501PC" (Japan) spectrophotometer in the standard 10 mm photometric quartz cuvette using THF solutions of the corresponding compounds with the concentrations of $1 \times 10^{-5}$ mol L$^{-1}$. All measurements were carried out at RT. The optical properties in films were studied using a SE2030-010-DUVN spectrophotometer with a wavelength range of 200–1100 nm.

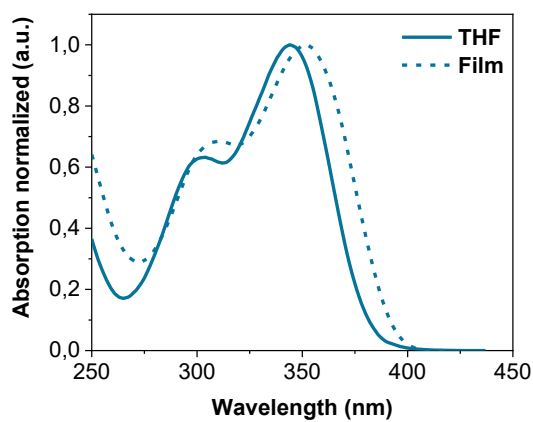

**Figure S4** - UV-Vis absorption spectra of **TPA-Py** in THF solution and thin films cast from THF

**CV**. Cyclic voltammetry (CV) measurements for TPA-Py solution were carried out with a three-electrode electrochemical cell in an inert atmosphere in an electrolyte solution, containing 0.1 M tetrabutylammonium hexafluorophosphate (Bu$_4$NPF$_6$) in THF using IPC-Pro M potentiostat. The scan rate was 200 mV s$^{-1}$. The glassy carbon electrode was used as the work electrode. The film was applied to a glassy carbon surface used as a working electrode by rubbing. A platinum plate placed in the cell served as the auxiliary electrode. Potentials were measured relative to a saturated calomel electrode (SCE). The highest occupied molecular orbital (HOMO) and the lowest unoccupied molecular orbital (LUMO) energy levels were calculated using the first formal oxidation and reduction potentials, respectively, obtained from CV experiments in acetonitrile according to following equations: LUMO = $-e(\varphi_{red}+4.40)$ (eV) and HOMO = $-e(\varphi_{ox}+4.40)$ (eV).

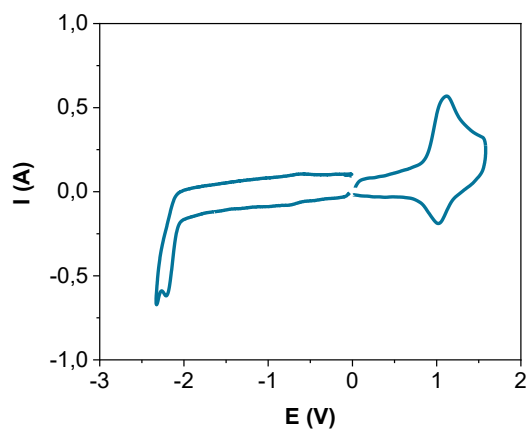

**Figure S5** - Cyclic voltammograms for solution of TPA-Py

**Quantum chemical calculations** were performed with ORCA program package (v. 6.1.0) [S3]. Calculation was done at PBE0/def2-QZVPD//$r^2$SCAN-3c level of theory. Electrostatic Potential surface was constructed with Multiwfn program (v. 3.8) [S4-S6], and plotted with VMD (v. 1.9.3) [S7]. Frontier orbitals were visualized with Iboview program [S8]

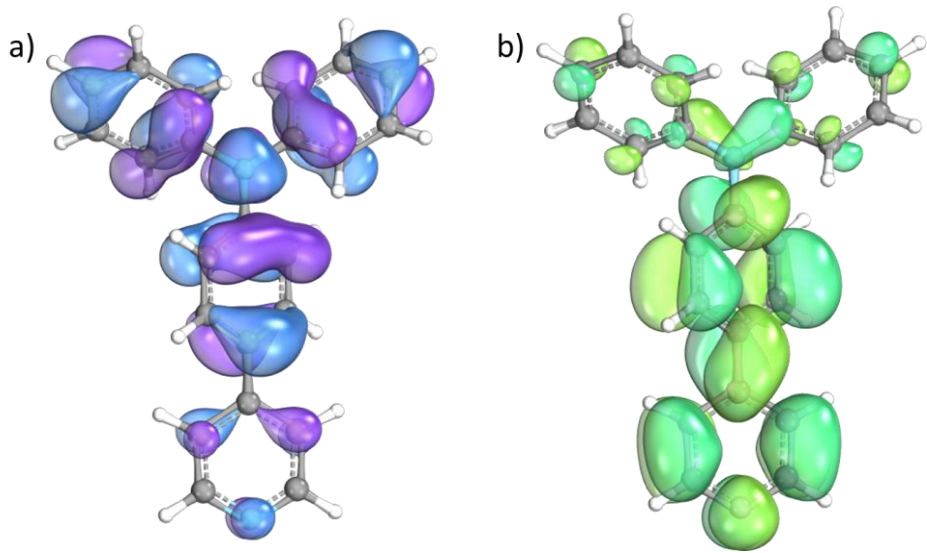

**Figure S6** – Calculated HOMO (a) and LUMO (b) molecular orbitals representation.

*Film characterization*

Absorbance spectra (**ABS**) of perovskite films were measured via SE2030-010-DUVN spectrophotometer with a wavelength range of 200–1100 nm. Static photoluminescence spectra (**PL**) were recorded on Agilent Cary Eclipse Fluorescence Spectrophotometer using 530 nm excitation wavelength. Time-resolved photoluminescence (**TRPL**) measurements were performed with time-correlated single-photon counter technique on Zolix OmniFluo-990 spectrofluorometer. The fluorescence was induced with 375 nm picosecond-pulsed laser (CNILaser MDL-PS-375). The signal acquisition was conducted until 20 000 counts. Atomic force microscopy (**AFM**) and Kelvin probe force microscopy (**KPFM**) measurements were performed in room ambient conditions using Ntegra AURA (NT-MDT) microscope. NSG10/Pt (Tipsnano) probes were used with tip curvature radius 30 nm. KPFM study utilized an amplitude modulation regime. AFM and KPFM images were obtained simultaneously by the two-pass method. Water contact angle was measured using KRÜSS EasyDrop DSA20., **SEM** images of perovskite film obtained with Vega3SB Tescan microscope. To investigate the **buried interface** of the perovskite film, a peeling procedure was employed. A UV-curable epoxy resin was deposited onto the perovskite surface and subsequently pressed with a glass substrate. After 25 seconds of UV irradiation, the perovskite film was detached from the original substrate.

*Device characterization*

Current density–voltage (**JV**) curves were measured in an ambient atmosphere by Keithley 2401 SMU with a settling time of $10^{-2}$ s and voltage step of 24 mV. The performance under 1 sun illumination conditions were measured with ABET technologies company (USA) Sun3000 solar simulator (1.5 AM G spectrum, 100 mW/cm$^{-2}$). The temperature of the devices was maintained at 25 °C. Solar simulator was calibrated to standard conditions with a certified

Si cell and an Ophir irradiance meter. The external quantum efficiency (**EQE**) spectra were measured using QEX10 solar cell quantum efficiency measurement system (PV Measurements Inc., USA) equipped with xenon arc lamp source and dual grating monochromator. Measurements were performed in DC mode in the 300–850 nm range at 10 nm step. The system was calibrated using the reference NIST traceable Si photodiode. The conformity of spectral response for the measured PSCs was calibrated with Si-solar cell and was compliant to the ASTM E 1021-06 standard. The difference between Jsc values gathered from JVs and those extracted from EQE measurements is mainly related to the different performances of the certified calibration cells used for adjustments to the standard conditions of illumination for the solar simulator and EQE system. Maximum power point tracking (**MPPT**) was performed with the following algorithm (software was developed in LabVIEW): forward JV scan, calculation of Pmax, Imax tracking at Vmax bias every 1 s. The transient photo-current (**TPC**) was measured with Tektronix TDS 3054C (oscilloscope) and Tektronix AFG 3252 (pulse generator), a square pulse of light was used to evaluate changes in the rise and fall profiles of the photocurrent over time. We used TDS-P001L4G05 STAR LED (540 nm) as a light source. Completed set up for characterization of PSCs was placed in a black box. A frequency of 10 kHz was applied to the LED. Specifically, we analyzed the rise and fall times (tr and tf) of the current response at amplitudes corresponding to 10% and 90% of signal saturation. To study **thermal stability**, the devices were maintained at a temperature of 85 °C and a relative humidity of no more than 60 %. JVs were measured every 20 hours until reaching 100 hours and every 40-70 hours afterward. Before measurements, the devices were lightsoaked for 1 hour. Admittance Spectroscopy **(AS)** measurements were performed using a LN120 liquid nitrogen cryostat (CryoTrade Engineering Ltd.) with a E4980A Precision LCR-meter (Keysight). Admittance of the samples was recorded as a function of temparature (within 180-350K range) and frequency (within 20Hz-1MHz range) (see **Fig. S12**). Temperature was swept at 1.5 K/min rate. Photo-induced Voltge Transient Spectroscopy (**PIVTS**) measurements used a 470 nm LED delivering ~250mW/cm$^2$ of optical power. Light pulse generation and recording of a $V_{OC}$ transients were done using B2902A Precision Source/Measure Unit (Keysight). Relaxations were recorded with 150ms time step in 180-350K temperature range (see **Fig. 4**).

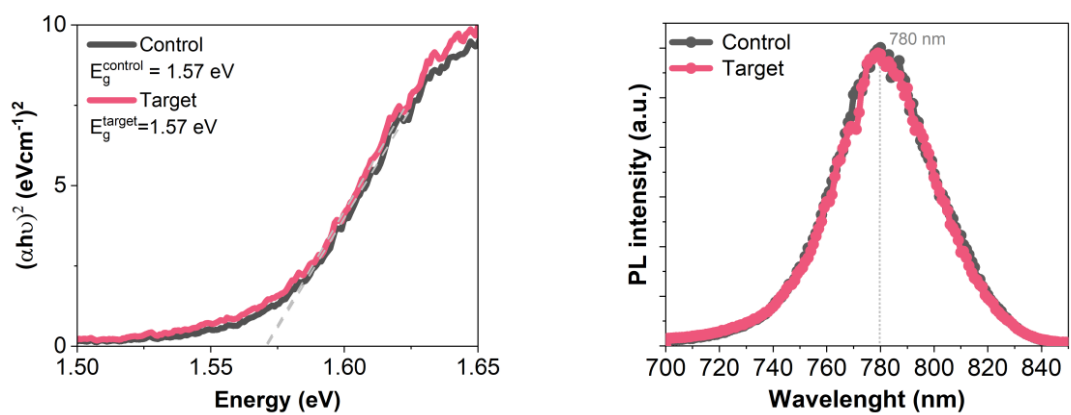

**Figure S7** – Tauc plot (a) and photoluminescence (b) spectra for control and target films

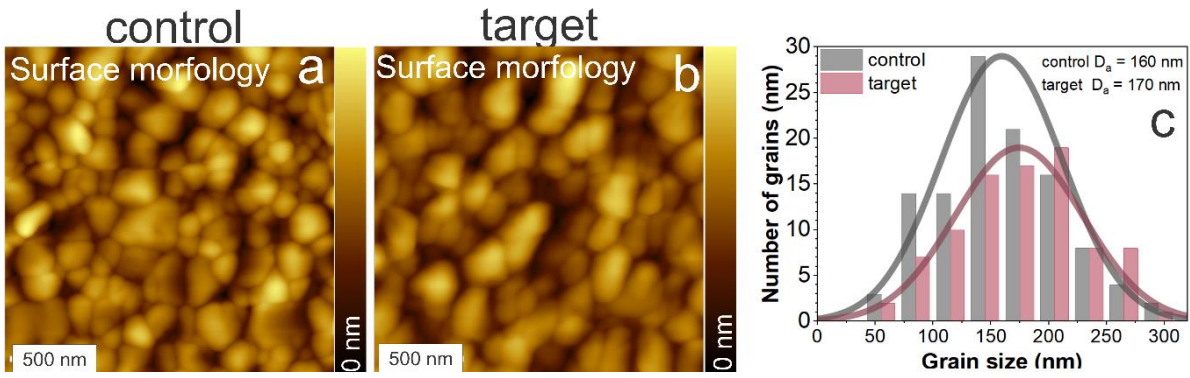

**Figure S8** – surface and buried interface morphology of control and target perovskite films

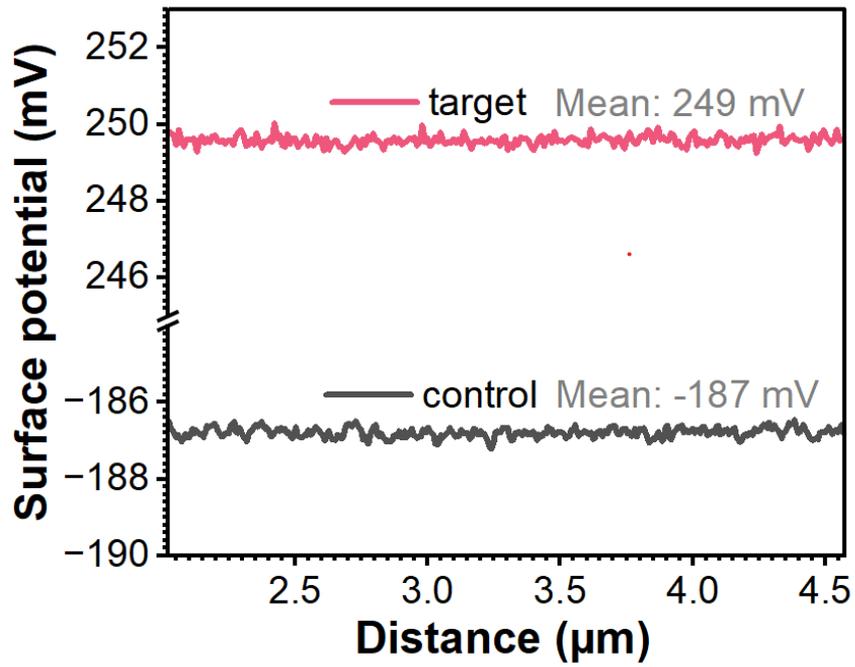

**Figure S9** – Surface potential distribution of control and target perovskite films

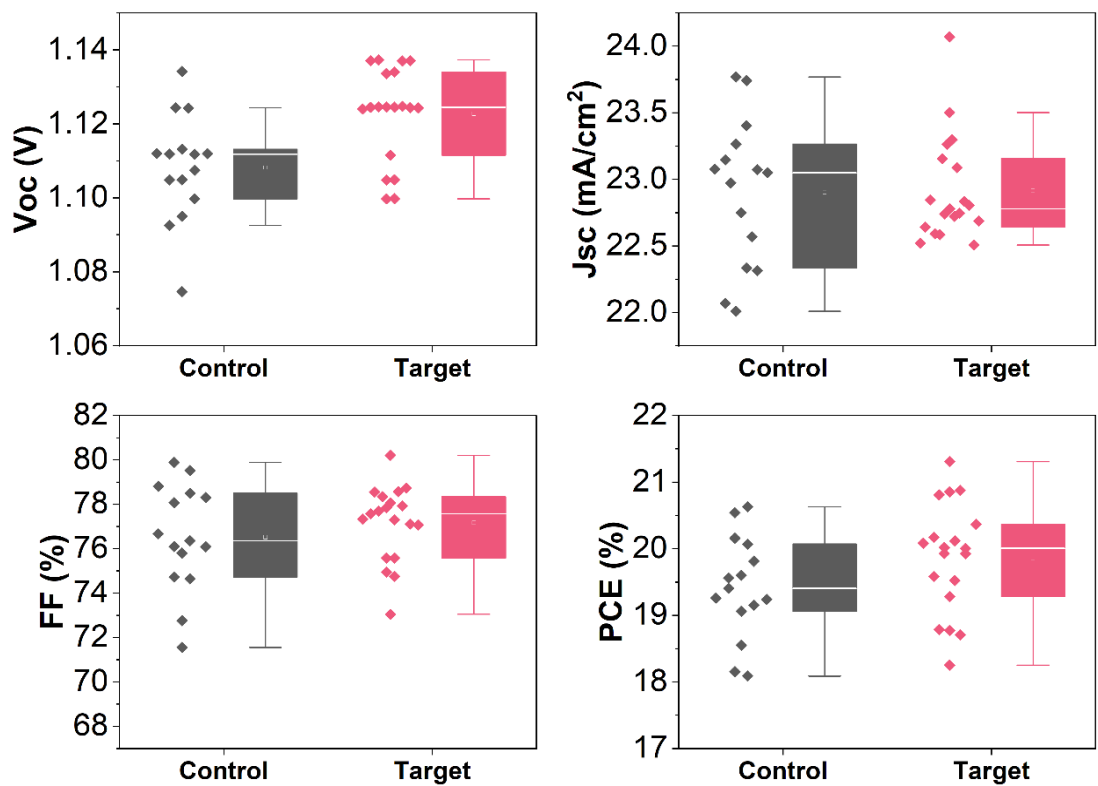

**Figure S10** – Box Charts JV statistics for the fabricated PSCs of Control and Target configurations

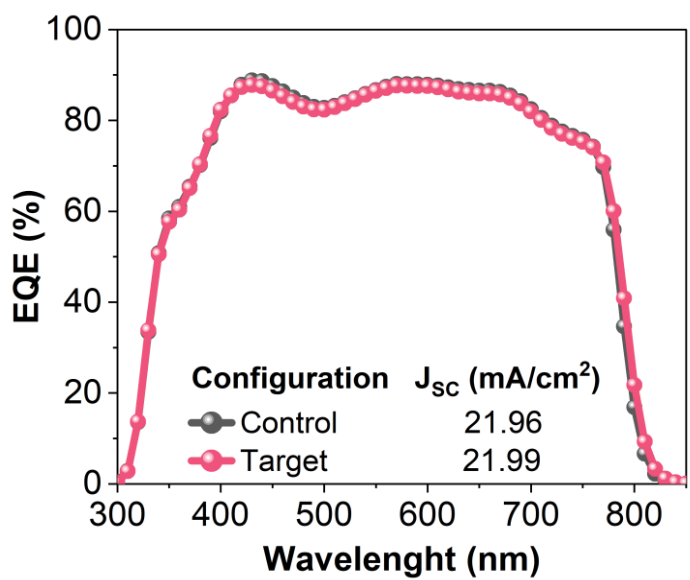

**Figure S11** – External quantum efficiency

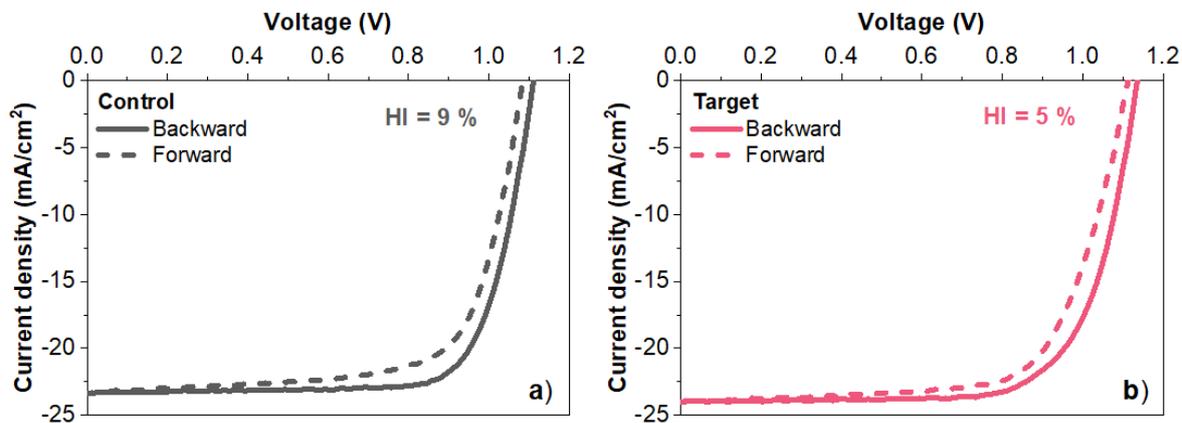

**Figure S12** – Forward and backward JVs of (a) Control and (b) Target devices with hysteresis index (HI) values

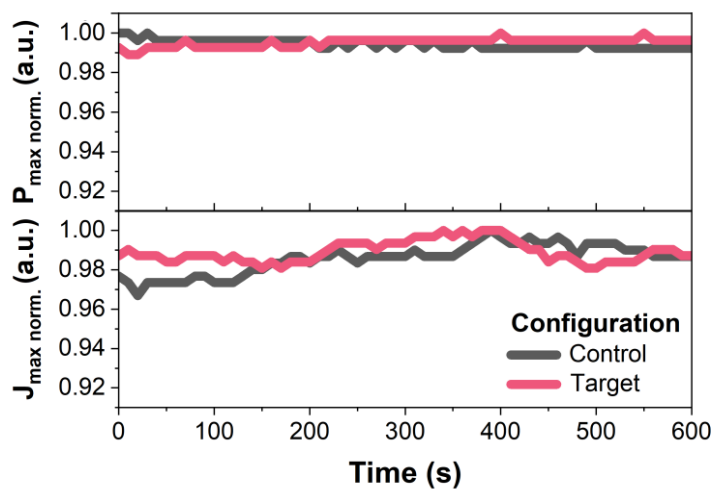

**Figure S13** – Maximum power point tracking measurements under standard illumination conditions

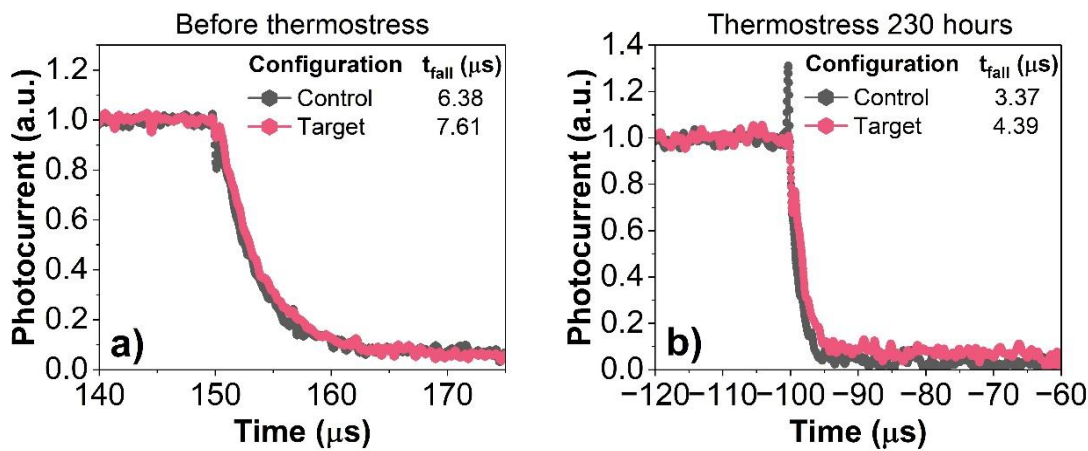

**Figure S14** – Transient photo-current measurements in fall mode for control and target devices before thermal stress (a) and after 230 hours of thermal stress (b)

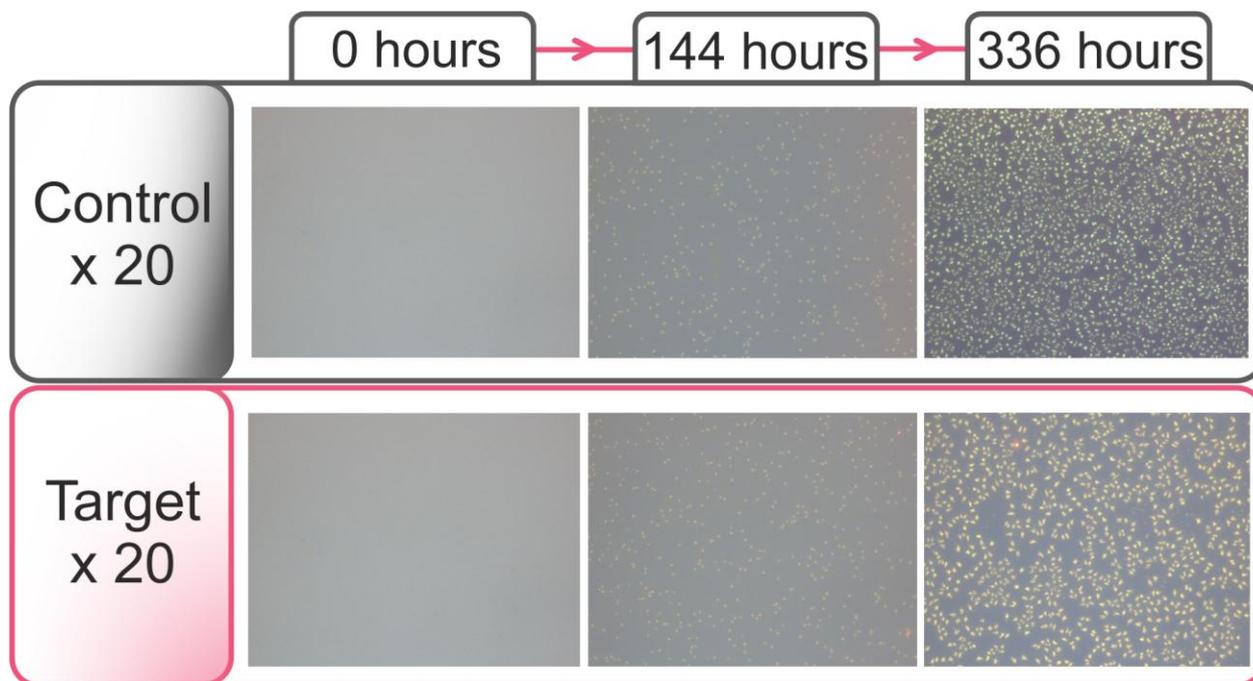

**Figure S15** – Optical microscopy of perovskite clusters for various periods of thermal stress

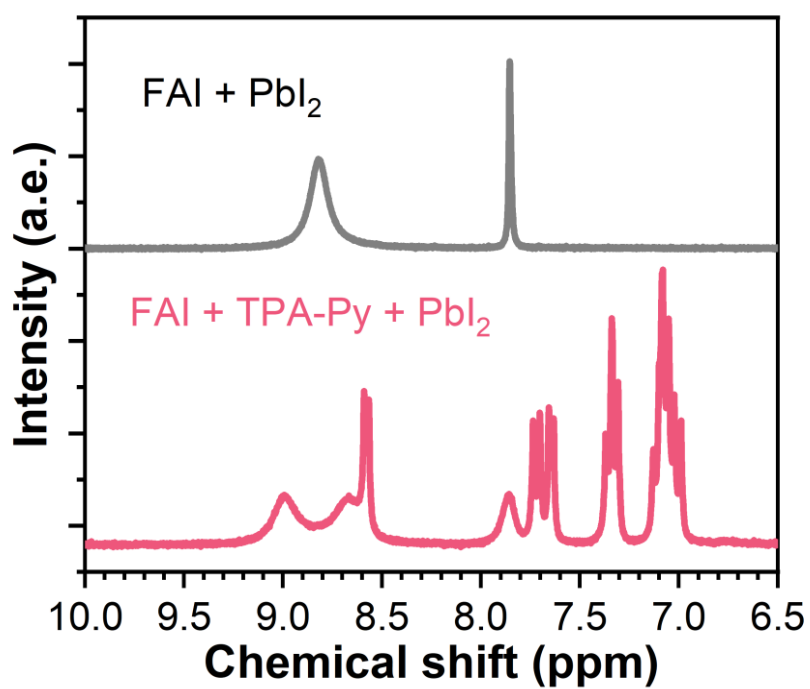

**Figure S16** – ¹H NMR spectra for the mixture of FAI and PbI$_2$ and for the mixture of FAI, PbI$_2$ and TPA-Py in DMSO-d6. The broadening and shift of the FAI signals may indicate the emergence of a coordination interaction between FAI and TPA-Py, corrected for the influence of DMSO-d6